\documentclass[aps,floats,twocolumn,showpacs]{revtex4-1}
\usepackage{amssymb}
\usepackage{amsmath}
\usepackage{epstopdf}
\usepackage{graphicx}

 \def\bc{\begin{center}}          \def\ec{\end{center}}
 \allowdisplaybreaks
\begin{document}
 \title{Electron trapping and acceleration by the plasma wakefield of a self-modulating proton beam}
 \author{K.V.Lotov}
 \affiliation{Budker Institute of Nuclear Physics SB RAS, 630090, Novosibirsk, Russia}
 \affiliation{Novosibirsk State University, 630090, Novosibirsk, Russia}
 \author{A.P.Sosedkin}
 \affiliation{Budker Institute of Nuclear Physics SB RAS, 630090, Novosibirsk, Russia}
 \affiliation{Novosibirsk State University, 630090, Novosibirsk, Russia}
 \author{A.V.Petrenko}
 \affiliation{Budker Institute of Nuclear Physics SB RAS, 630090, Novosibirsk, Russia}
 \affiliation{CERN, CH-1211 Geneva 23, Switzerland}
 \author{L.D.Amorim, J.Vieira, R.A.Fonseca, L.O.Silva}
 \affiliation{Instituto de Plasmas e Fus\~ao Nuclear, Instituto Superior T\'ecnico, Universidade de Lisboa, 1049-001, Lisbon, Portugal}
 \author{E.Gschwendtner}
 \affiliation{CERN, CH-1211 Geneva 23, Switzerland}
 \author{P.Muggli}
 \affiliation{Max Planck Institute for Physics, 80805 Munich, Germany}
 \date{\today}
 \begin{abstract}
It is shown that co-linear injection of electrons or positrons into the wakefield of the self-modulating
particle beam is possible and ensures high energy gain. The witness beam must co-propagate with the tail part of the driver, since the plasma wave phase velocity there can exceed the light velocity, which is necessary for efficient acceleration. If the witness beam is many wakefield periods long, then the trapped charge is limited by beam loading effects. The initial trapping is better for positrons, but at the acceleration stage a considerable fraction of positrons is lost from the wave. For efficient trapping of electrons, the plasma boundary must be sharp, with the density transition region shorter than several centimeters. Positrons are not susceptible to the initial plasma density gradient.
  \end{abstract}
 \maketitle

\section{Introduction}

Proton driven plasma wakefield acceleration (PDPWA) is now actively studied as a possible path to future high energy colliders \cite{AWAKE}. The interest is motivated by the ability of plasmas to support extremely strong electric fields \cite{RMP81-1229} and by the availability of proton beams carrying tens of kilojoules of energy in a single bunch \cite{PRD86-010001}. The high energy content of proton beams makes it possible to accelerate multi-nanocoulomb electron bunches to sub-teraelectronvolt energies and beyond in a single plasma stage \cite{NatPhys9-363,PRST-AB13-041301}, which is the main advantage of PDPWA over other plasma wakefield acceleration schemes.

The initial proposal of PDPWA \cite{NatPhys9-363} assumed longitudinal compression of the proton bunch to a sub-millimeter length, which is difficult to realize \cite{PAC09-4542,PPCF53-014003,AIP1229-510,IPAC10-4395}. The effect of beam self-modulation in the plasma \cite{EPAC98-806,PPCF53-014003,PRL104-255003}, however, makes proof-of-principle experiments on PDPWA possible without costly conditioning of the proton beam prior to the plasma. The experiment named AWAKE thus started at CERN \cite{AWAKE,IPAC13-1179,TDR,NIMA-740-48}, as well as several supporting experiments with electron beams \cite{PRL112-045001,NIMA-740-74,AIP1229-467,PoP19-063105,IPAC13-1235}.

Injection of electrons into the wakefield of a self-modulating beam turned out to be a nontrivial task. During development of the self-modulation instability, the phase velocity of the wakefield is substantially lower than the light velocity $c$, as was pointed out in Refs.\cite{PRL107-145003,PRL107-145002}. It was predicted that the electron energy gain in a PDPWA driven by a self-modulated beam in a uniform plasma will be severely limited by dephasing \cite{PRL107-145002}, and tapering the plasma density was proposed to overcome the dephasing limit \cite{PRL107-145003,PoP19-010703}. Another possible way to high electron energies involves side injection of electrons into the plasma wave at the stage of fully developed self-modulation \cite{PRL107-145003,JPP78-455}. Although the side injection is expected to produce good energy spectra of accelerated electrons \cite{AWAKE,IPAC13-1238,TDR}, its implementation presents some technical difficulties. The parameter window for good trapping is rather narrow, and the low energy electron beam must be first transported through the highly uniform vapor \cite{PoP20-013102,NIMA-740-197} for several meters and only then injected into a certain region at a certain angle.

\begin{figure*}[t]
\bc\includegraphics[width=473bp]{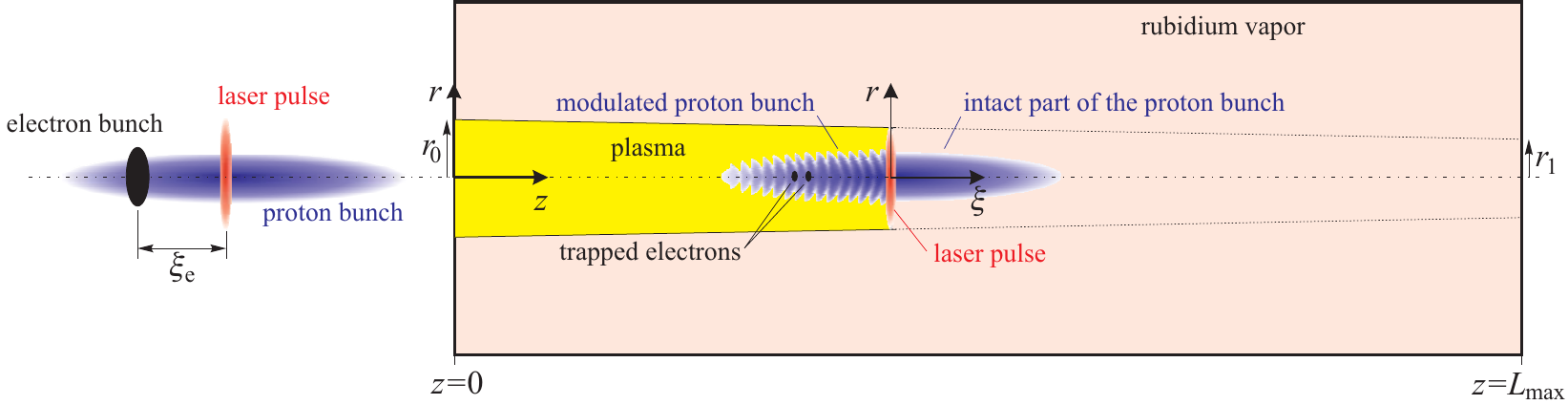} \ec
\caption{Geometry of the problem (not to scale). The beams are shown at two times.}\label{f1-setup}
\end{figure*}
\begin{table}[b]
 \caption{ Baseline AWAKE parameters and notation.}\label{t1}
 \bc\begin{tabular}{ll}\hline
  Parameter, notation & Value \\ \hline
  Plasma density, $n_0$ & $7 \times 10^{14}\,\text{cm}^{-3}$ \\
  Plasma length, $L_\text{max}$ & 10\,m \\
  Atomic weight of plasma ions, $M_i$ & 85.5 \\
  Plasma skin depth, $c/\omega_p \equiv k_p^{-1}$, & 0.2\,mm \\
  Initial plasma radius, $r_0$, & 1.5\,mm \\
  Final plasma radius, $r_1$, & 1\,mm \\
  Wavebreaking field, $E_0=mc\omega_p/e$, & 2.54\,GV/m \\
  Proton bunch population, $N_b$ & $3\times 10^{11}$ \\
  Proton bunch length, $\sigma_{zb}$ & 12\,cm \\
  Proton bunch radius, $\sigma_{rb}$ & 0.2\,mm \\
  Proton bunch energy, $W_b$ & 400\,GeV \\
  Proton bunch energy spread, $\delta W_b$ & 0.35\% \\
  Proton bunch normalized emittance, $\epsilon_{nb}$ & 3.6\,mm\,mrad \\
  Proton bunch maximum density, $n_{b0}$ & $4\times 10^{12}\,\text{cm}^{-3}$ \\
  Electron bunch population, $N_e$ & $1.25\times 10^9$ \\
  Electron bunch length, $\sigma_{ze}$ & 1.2\,mm \\
  Electron bunch radius, $\sigma_{re}$ & 0.25\,mm \\
  Electron bunch energy, $W_e$ & 16\,MeV \\
  Electron bunch energy spread, $\delta W_e$ & 0.5\% \\
  Electron bunch normalized emittance, $\epsilon_{ne}$ & 2\,mm\,mrad \\
  Electron bunch delay, $\xi_e$ & 16.4\,cm\\ \hline
 \end{tabular}\ec
\end{table}
In this paper we demonstrate that the on-axis injection of electrons into the wakefield of the self-modulating particle beam can ensure good trapping and acceleration even in the uniform plasma. By the on-axis injection we mean propagation of both electron and proton beams along the same line starting from entrance to the plasma. The novel effect that enables the better performance is the appearance of a supraluminal wave at the stage of developed self-modulation. As reference case we take the latest AWAKE baseline parameters (Table~\ref{t1}).

The process is studied numerically with three codes: fluid \cite{PoP5-785,LCODE} and particle-in-cell \cite{LCODE,PRST-AB6-061301,IPAC13-1238} versions of 2d3v quasi-static code LCODE and with cylindrically symmetric (2d3v) particle-in-cell code OSIRIS \cite{osiris}. By performing 2D cylindrically symmetric simulations we preclude the physics associated with the hosing instability, which can lead to beam breakup~\cite{PRL104-255003,PRE86-026402,PoP20-056704}. It has been shown, however, that the hosing instability can be suppressed after the saturation of the self-modulation instability if wakefield excitation is in the linear regime~\cite{PRL112-205001}. Since the baseline AWAKE variant will lead to plasma wakefields excited in the linear regime, cylindrically symmetric simulations are well suited for our research. Unless stated otherwise, figures are produced with the particle-in-cell LCODE. The fluid code is used for trapping studies, as it produces less noisy results at the initial stage of beam evolution. The main findings are also demonstrated and cross-checked in OSIRIS simulations. In trapping studies we analyze both electrons and positrons, as comparison of the two gives a better insight into the trapping mechanism.

The simulated setup is shown in Fig.\,\ref{f1-setup}. Three superimposed beams (proton, electron, and laser) propagate collinearly through the volume filled with a uniform rubidium vapor. The short laser pulse singly ionizes the vapor and creates the plasma of radius $r_p(z)$ that varies linearly from $r_0$ at the plasma entrance to $r_1 < r_0$ at the plasma exit. The leading half of the proton bunch propagates in the neutral gas and does not contribute to wakefield excitation. The rear half of the proton beam undergoes self-modulation. The self-modulation instability is seeded by the instant onset of the plasma, which acts as if the bunch has a sharp leading edge. We also note that the atomic weight of the plasma ions is sufficiently large to avoid deleterious effects associated with the plasma ion motion~\cite{PoP21-056705}. The electron bunch is delayed with respect to the laser pulse by the distance $\xi_e$. We use cylindrical coordinates $(r, \varphi, z)$ with the $z$-axis as the direction of beam propagation and the co-moving coordinate $\xi = z-ct$ measured from the laser pulse.
Focusing and accelerating properties of the plasma wave are most conveniently characterized by the quantity
\begin{equation}\label{e1}
    \Phi (r,\xi,z) = \frac{\omega_p}{E_0} \int_{-\infty}^{\xi/c} E_z(r, z, \tau) \, d \tau,
\end{equation}
where $E_z$ is the longitudinal electric field. If the time scale of beam evolution is much longer than the wave period, then  (\ref{e1}) is close to the dimensionless wakefield potential, so we refer to it as the wakefield potential also.

\begin{figure*}[t]
\bc\includegraphics[width=449bp]{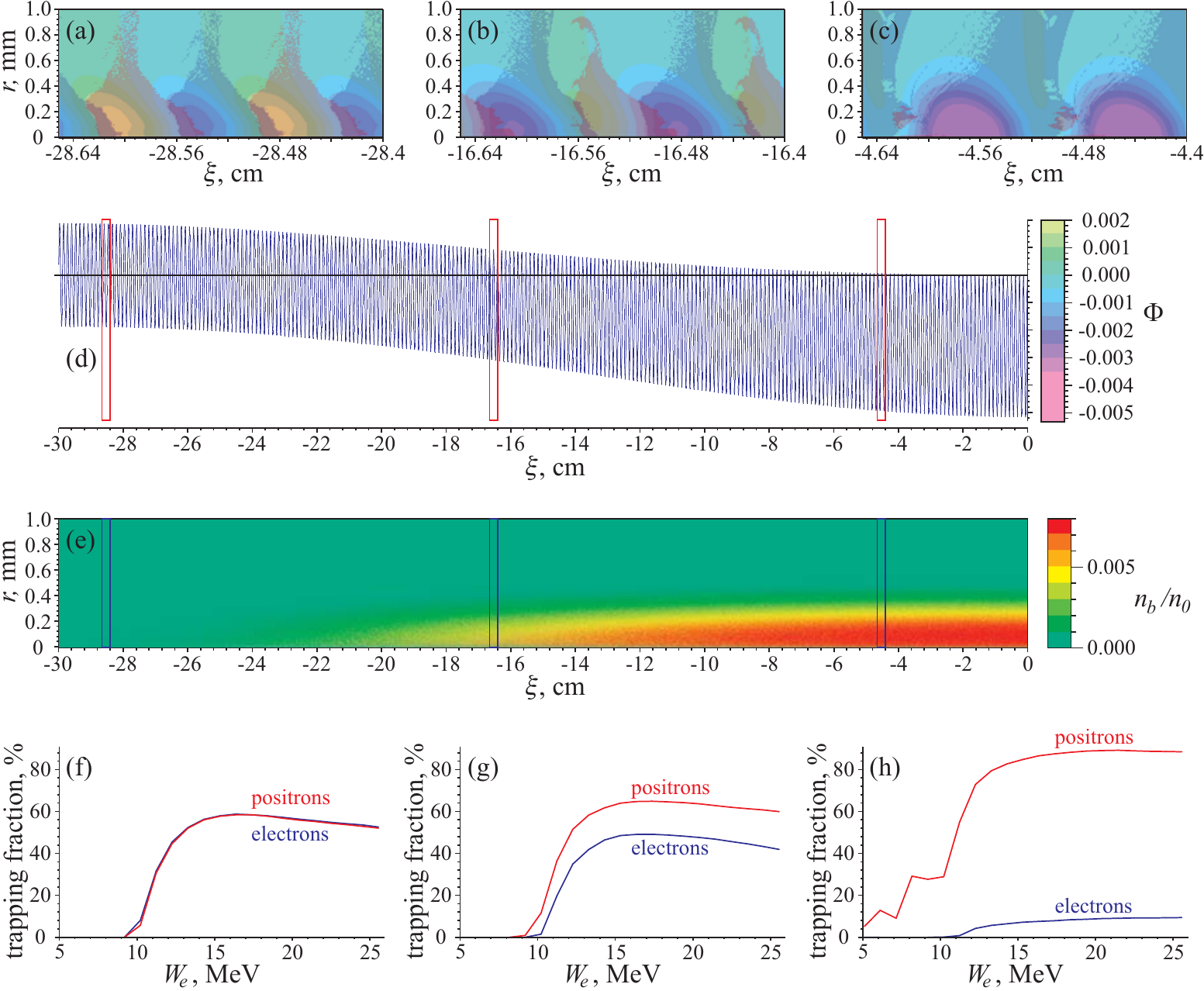} \ec
\caption{ (a)-(c) Acceptance of the plasma wave for positrons (blue dots) and electrons (red dots) plotted over the potential map at three locations along the proton bunch; (d) the wakefield potential on the axis; (e) the corresponding map of the proton beam density; (f)-(h) dependence of the trapping fraction on the electron or positron beam energy for the selected locations. The beams propagate to the right. The locations of the zoomed in regions (a)-(c) are shown in (d) and (e) by narrow rectangles; the color map for the potential is shown in (d). Simulations are made with the fluid code LCODE.}\label{f2-trapping}
\end{figure*}
In Section~\ref{s2} we study trapping of test particles, that is we exclude the back action of trapped particles on the wakefield to describe the trapping process in the cleanest form. In Section~\ref{s3} we focus on subsequent acceleration of test particles. Then we discuss the effect of beam loading in Section~\ref{s4} and the effect of smooth plasma boundary in Section~\ref{s5}.

\section{Trapping of test particles} \label{s2}

First we note that separation of injected particles into trapped and untrapped fractions occurs at the very beginning of the plasma, before the drive beam has time to self-modulate. Indeed, the depth of the transverse potential well initially formed by the seed perturbation is \cite{PoP18-103101}
\begin{equation}\label{e2}
    W_f \sim mc^2 \frac{n_{b0}}{4 n_0} \approx 1.5 \times 10^{-3} mc^2.
\end{equation}
The initial energy of transverse electron motion can be estimated as \cite{PoP18-103101}
\begin{equation}\label{e3}
    W_{tr} \sim mc^2 \frac{\epsilon_{ne}^2}{2 \gamma_e \sigma_{re}^2} \approx 7 \times 10^{-7} mc^2,
\end{equation}
where $\gamma_e = W_e/(mc^2)$. Thus, for any proton beam of interest and high quality electron bunches, the initial transverse velocity of electrons can be safely neglected. The longitudinal velocity could have an effect on trapping, but, as we show later, this effect can be minimized by matching the electron velocity and the phase velocity of the wave. Whether a particle is trapped or not is thus determined by the particle location in the initial wakefield potential.

\begin{figure*}[t]
\bc\includegraphics[width=464bp]{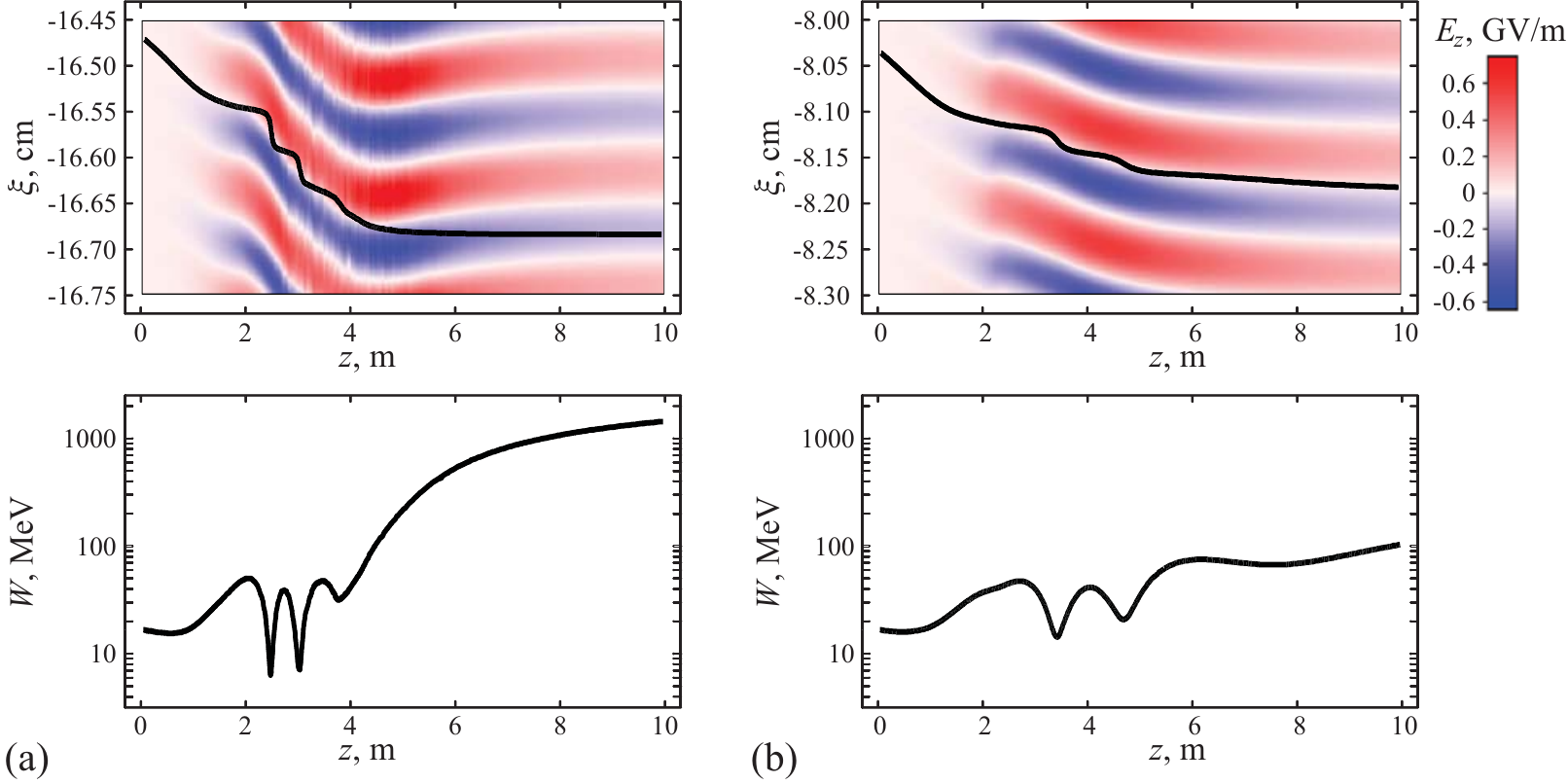} \ec
 \vspace*{-5mm}
\caption{ The co-moving coordinate $\xi$ (top) and the energy (bottom) versus the propagation distance for two typical test electrons injected with different delays with respect to the laser pulse. The top plots also show the color map of the on-axis electric field $E_z$ in the vicinity of the electron.} \label{f3-phases}
\end{figure*}
Simulations confirm this inference. In Fig.\,\ref{f2-trapping}(a-c) we show initial positions of subsequently trapped electrons or positrons and the potential profile in these regions. To be exact, by trapping we mean that the particle remains at $r < 3 c/\omega_p$ after 1 meter of propagation in the plasma. We can see that trapped particles initially reside in potential wells or are separated radially from outer regions by potential crests.

There is also a clear trapping asymmetry between electrons and positrons, which is stronger the closer the injected bunch to the center of the proton bunch is. The asymmetry is due to incomplete neutralization of the proton beam current. It has long been known that there is a complete local neutralization of the beam charge by the dense plasma, while the current neutralization is essentially non-local if the beam radius is smaller than or on the order of $c/\omega_p$ (see, e.g., Refs.~\cite{PP15-429,PoP3-2753}). This gives rise to the well known plasma lens effect \cite{PAcc20-171}, that is strong focusing of a charged particle beam by its own incompletely neutralized magnetic field. The wakefield potential is thus the sum of two terms. One term is due to the seed perturbation; it oscillates with the plasma frequency, and oscillation amplitude is proportional to the proton bunch density in the central cross-section. Another term is due to the plasma lens effect; it smoothly varies along the beam, and its value is proportional to the proton current at the considered cross-section. Both terms have the same radial dependence, as follows from the linear wakefield theory \cite{PF30-252}, and cancel at $\xi=0$. In the central part of the proton beam, the total potential is thus negative almost everywhere, which means focusing for positrons and defocusing for electrons [Fig.\,\ref{f2-trapping}(c,d)]. As the beam density decreases towards the beam tail, the lens effect vanishes, and trapping areas for electrons and positrons gradually equalize [Fig.\,\ref{f2-trapping}(a-e)].

To study the energy dependence of trapping, we introduce the trapping fraction as the number ratio of trapped to injected test particles [Fig.\,\ref{f2-trapping}(f-h)]. In our case test particles uniformly cover a rectangle two wave periods in length and $1.5\, c/\omega_p$ in radius. Note that this quantity is not a quantitative measure of trapping for a real beam, as the trapped charge depends on the beam density distribution, so only qualitative inferences can be made from Fig.\,\ref{f2-trapping}(f-h). We see that there is a cut-off energy below which trapping is not possible in most cross-sections. The maximum trapping fraction is observed at energies for which the velocity of injected particles is close to the phase velocity of the wave at the self-modulation stage. Higher energy particles are also well trapped.

\section{Acceleration of test particles} \label{s3}

Once a particle is trapped by the wakefield, it follows the potential well (Fig.\,\ref{f3-phases}). During the development of the self-modulation, the particle makes several longitudinal oscillations and many transverse oscillations in the potential well. The particle energy also oscillates around its initial value. After the proton beam is fully micro-bunched, trapped particles are either accelerated to high energies or not depending on their location with respect to the seed laser pulse. At large $|\xi|$, the wave phase velocity becomes greater than the speed of light, and the trapped particles (which cannot catch up with the bottom of the potential well) shift to regions of strong accelerating field [Fig.\,\ref{f3-phases}(a)]. At small $|\xi|$, the wave remains subluminal, and no continuous acceleration occurs [Fig.\,\ref{f3-phases}(b)] resulting in $W<100$\,MeV versus $W>1$\,GeV in the baseline case of large $|\xi|\sim 600 c/\omega_p$.

There are two reasons for the appearance of the supraluminal wakefield. The first one is related to the nonlinear elongation of the wave period at high wakefield amplitudes \cite{PoP20-083119}. As the wave amplitude reduces after peaking at $z \sim 4$\,m, the wavelength returns to its low-amplitude value $2 \pi c / \omega_p$, and the wave at the driver tail moves forward with respect to the driver. The second reason comes from the relative positioning of the wake and proton micro-bunches formed by the self-modulation instability. The bunches are delayed with respect to points of maximum decelerating field (see Fig.\,3a of Ref.\,\cite{PoP18-024501}). Consequently, each micro-bunch contributing the wakefield adds some backward shift to the wave. Once some micro-bunches are destroyed at the late stage of propagation \cite{PoP18-024501}, the wave shifts forward in the co-moving frame.

\begin{figure}[t]
\bc\includegraphics[width=224bp]{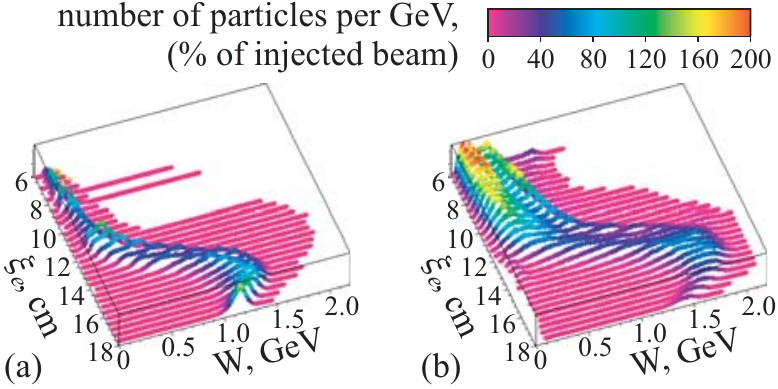} \ec
 \vspace*{-5mm}
\caption{ Final energy spectra for (a) electrons and (b) positrons as a function of injection delay $\xi_e$ with no beam loading effect taken into account.}\label{f4-array}
\end{figure}

To obtain a general grasp of the wave acceleration capability, we compare energy spectra of test beams injected at different delays with respect to the seed pulse (Fig.\,\ref{f4-array}). Each thick line in Fig.\,\ref{f4-array} is the final energy spectrum of a Gaussian-like electron or positron beam with all the parameters taken from Table~\ref{t1} except $\xi_e$ which is varied. The spectra are normalized to the number of particles in the injected beam, so the area under the curve is the beam trapped fraction in percent. We see that for this particular driver, the acceleration is possible for $\xi_e \gtrsim 12$\,cm, and the optimum is observed at $\xi_e \approx 16$\,cm. There is no much difference between acceleration of test electrons and positrons, though the higher efficiency of positron trapping at small $\xi_e$ translates to a larger number of weakly accelerated positrons.

\begin{figure}[t]
\bc\includegraphics[width=235bp]{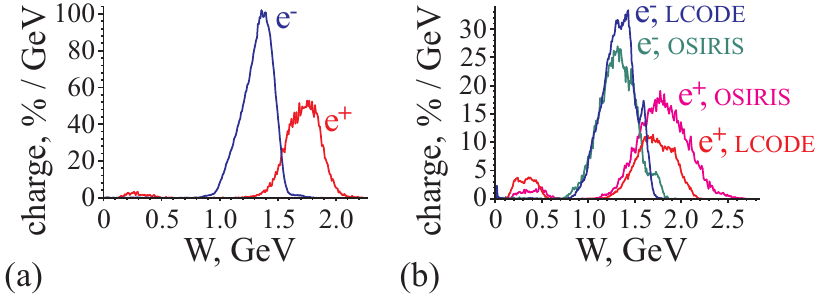} \ec
 \vspace*{-5mm}
\caption{ Final energy spectra of electron and positron bunches injected at the nominal delay $\xi_e = 16.4$\,cm without (a) and with (b) beam loading.}\label{f5-spectra}
\end{figure}
The final energy spectra for the nominal delay $\xi_e = 16.4$\,cm are shown in Fig.\,\ref{f5-spectra}(a). The fraction of accelerated particles is 31\% for electrons and 26\% for positrons. By comparison, the trapped fractions observed at $z=1$\,m are 32\% for electrons and 49\% for positrons. Apparently, this asymmetry is due to the above-mentioned plasma lens effect. Trapped electrons initially reside near the bottom of the potential well, and remain trapped as the potential well evolves. In contrast, positrons initially fill a wider area and are partially lost as the well changes its speed or shape.
\begin{figure}[b]
\bc\includegraphics[width=226bp]{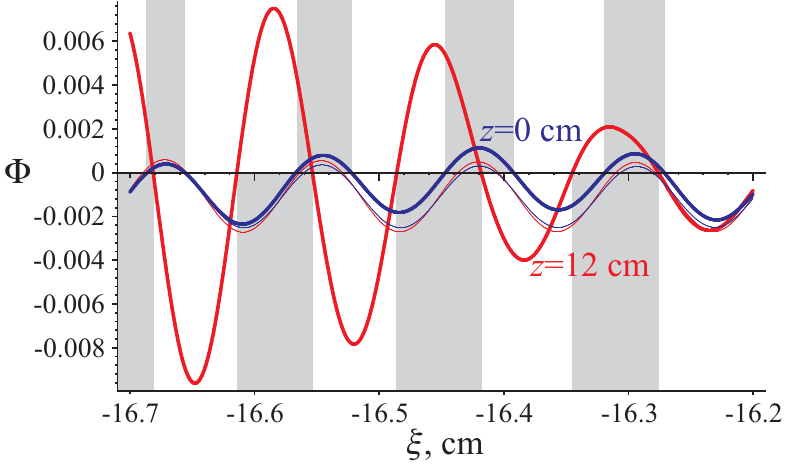} \ec
 \vspace*{-5mm}
\caption{ The on-axis wakefield potential at the very entrance to the plasma (blue) and at $z=12$\,cm (red). The two thin lines are the wakefield potential of the driver only; the two thick lines are the wakefields modified by the electron beam of population $N_e = 1.25\times 10^9$. Shading of the upper (lower) area shows the focusing regions for $z=0$\,cm ($z=12$\,cm). }\label{f6-loading}
\end{figure}

\section{Beam loading effect} \label{s4}

Taking into account the effect of the trapped charge on the wakefield, i.e., the beam loading, considerably reduces the number of accelerated particles, but has a small effect on the shape of the energy spectrum [Fig.\,\ref{f5-spectra}(b)]. From LCODE simulations, the accelerated fraction is 14.3\% for electrons and 6.8\% for positrons. From OSIRIS simulations, these numbers are 12.6\% for electrons and 11.7\% for positrons. The reason for the smaller numbers is that the wakefield of the particles trapped earlier (at smaller $|\xi|$) acts as a defocusing force. Though having a relatively small total charge (0.8\% of that in the drive beam), the injected beam is short, and therefore has a high peak current of 20\,A, which is comparable to the peak proton beam current (50\,A at $\xi=0$, 20\,A at $\xi_e$). The effect of wakefield distortion by the trapped beam can thus be very important.

\begin{figure}[t]
\bc\includegraphics[width=188bp]{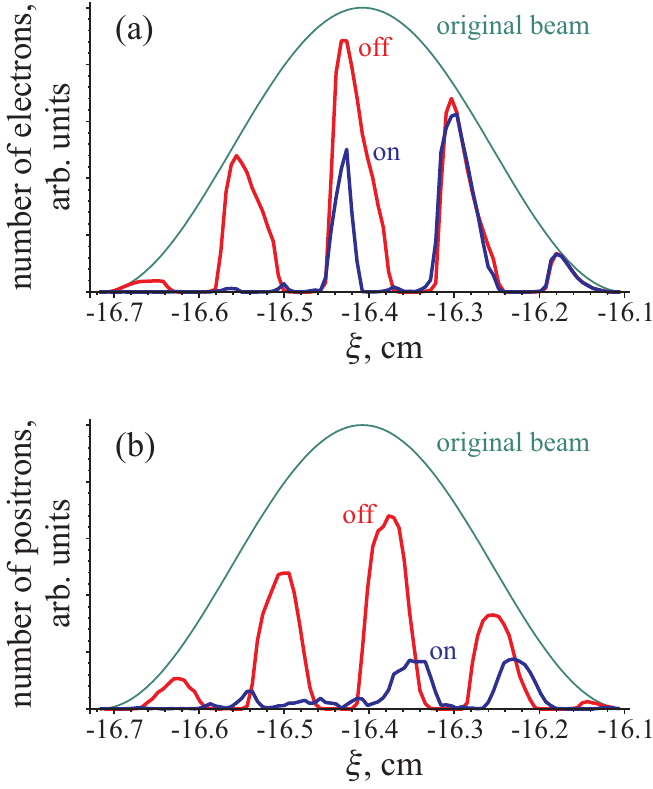} \ec
 \vspace*{-5mm}
\caption{ Number of electrons (a) and positrons (b) trapped at different cross-sections of the injected beam with the effect of beam loading on and off. The upper thin curves show the population of the original beams.}\label{f7-lock}
\end{figure}
\begin{figure}[t]
\bc\includegraphics[width=211bp]{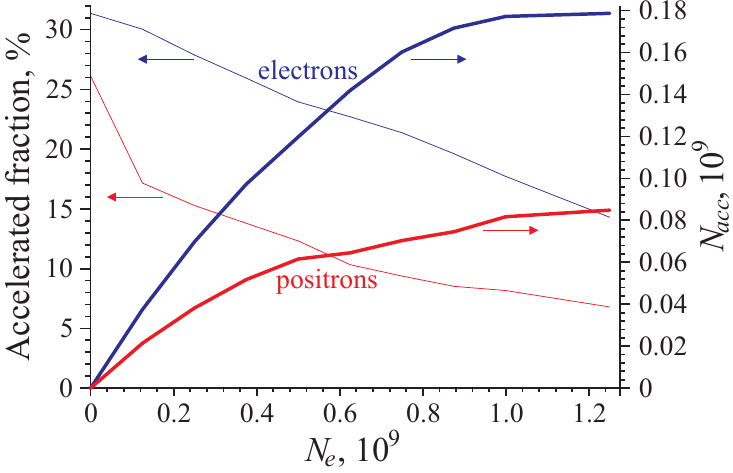} \ec
 \vspace*{-5mm}
\caption{ Fraction of accelerated particles (thin lines, left scale) and total number of accelerated particles $N_{acc}$ (thick lines, right scale) versus the number of injected particles $N_e$ for electron and positron beams.}\label{f8-fraction}
\end{figure}
We illustrate the effect in the electron beam case (Fig.\,\ref{f6-loading}).
At the very beginning of the interaction (at $z=0$), the electron beam is smooth, and its wakefield only contributes (favorably) to the plasma lens effect. In Fig.\,\ref{f6-loading} this is seen as a small upward shift of the potential (thick blue curve) with respect to the unloaded case (thin curve). Initial trapping of electrons proceeds in accordance with the initial potential shape, and the trapped electrons are located at cross-sections marked in grey in the upper part of Fig.\,\ref{f6-loading}. Once trapped, electrons form short micro-bunches, their wakefield strongly increases (thick red curve) and for some time dominates over the wakefield of the proton beam. During this period, the location of focusing areas changes, and only those electrons survive which are at the cross-sections marked in grey in the lower part of Fig.\,\ref{f6-loading}. As we see, at the rear part of the electron beam ($|\xi|>16.65$\,cm) the two grey areas almost do not overlap at all, which results is loss of particles [Fig.\,\ref{f7-lock}(a)]. For positrons, the picture is qualitatively the same [Fig.\,\ref{f7-lock}(b)]. Curiously, with the account of beam loading the number of accelerated positrons at some cross-sections is higher because of the plasma lensing, so the back effect of the trapped charge is not necessarily negative. The observed difference between the two codes comes from that the trapped fraction (unlike the final energy) is determined by the interplay of low-amplitude wakes which suffer from noise in particle-in-cell simulations.

``Closing'' the wakefield by the trapped charge is quantitatively characterized in Fig.\,\ref{f8-fraction}. As the charge of the injected beam grows, its accelerated fraction decreases, and the total accelerated charge comes to saturation. Perhaps the saturation effect can be avoided with shorter injected beams which cover one wakefield period only.

\section{Entry into the plasma} \label{s5}

At the beginning of the plasma section (at some transition region), the plasma density smoothly increases from zero to the nominal value, and the wakefield phase is rapidly changing in the vicinity of the witness beam. The consequences of that density variation depend on the ratio of three distances. The first one is the length of the transition region $L_0$. The second one is the defocusing length $L_d$ which characterizes radial scattering of witness particles by an unfavorable wakefield phase. This distance is determined by the radial force exerted on an axially moving electron by the driver wakefield \cite{PF30-252}:
\begin{multline}\label{e4}
    F_\perp (r,\xi) = 4 \pi e^2 k_p n_{b0} \\ \times \int_\xi^0 d\xi' e^{-{\xi'}^2/(2 \sigma_{zb}^2)} \sin \bigl( k_p (\xi'-\xi) \bigr) \\
\times \left(  \int_0^r dr' r' I_1(k_p r') K_1(k_p r) \frac{\partial e^{-{r'}^2/(2 \sigma_{rb}^2)}}{\partial r'} \right. \\
 + \left. \int_r^\infty dr' r' I_1(k_p r) K_1(k_p r') \frac{\partial e^{-{r'}^2/(2 \sigma_{rb}^2)}}{\partial r'} \right),
\end{multline}
where $I_1$ and $K_1$ are modified Bessel functions. For $k_p \sigma_{rb} = 1$ and the near-axis region ($k_p r \ll 1$), the sum in parentheses is approximately $0.27 r$. The integral over $\xi'$ can be transformed to
\begin{multline}\label{e5}
    k_p e^{-\xi^2/(2 \sigma_{zb}^2)} - k_p \cos k_p \xi \\
    + k_p \int_\xi^0 d\xi' \cos \bigl( k_p (\xi'-\xi) \bigr) \frac{\partial e^{-{\xi'}^2/(2 \sigma_{zb}^2)}}{\partial \xi'}.
\end{multline}
If the beam is long ($k_p \sigma_{zb} \gg 1$), the integral in (\ref{e5}) is small and can be neglected. The transverse force on a near-axis electron is thus
\begin{equation}\label{e6}
    F_\perp (r,\xi) = 4 \pi e^2 A_\perp n_{b0} r \left( e^{-\xi^2/(2 \sigma_{zb}^2)} - \cos (k_p \xi) \right)
\end{equation}
with $A_\perp \approx 0.27$. The second term in (\ref{e6}) is the seed perturbation produced by the ionization front; the first term (always positive) accounts for defocusing by the uncompensated current of the proton beam. The typical defocusing distance can be thus estimated as
\begin{equation}\label{e7}
    L_d \sim c \sqrt{\frac{\gamma_e m}{4 \pi e^2 A_\perp n_{b0}}} \approx 3\,\text{cm}.
\end{equation}
Note that this distance does not depend on the plasma density.

The third important length, $L_n$, is the distance at which the wakefield experienced by a witness particle changes its phase by $\pi$ because of the plasma density change. Assume for simplicity that the local plasma density $n_p$ is growing linearly:
\begin{equation}\label{e8}
    n_p (z) = n_0 z / L_0, \qquad z < L_0.
\end{equation}
The distance $\xi_e$ between the seed pulse and the witness bunch corresponds to the phase advance
\begin{equation}\label{e9}
    \phi = \frac{\xi_e}{c} \sqrt{\frac{4 \pi n_p (z) e^2}{m}},
\end{equation}
whence
\begin{equation}\label{e10}
    L_n = \pi \left(  \frac{\partial \phi}{\partial z}  \right)^{-1} = \frac{2 \pi c \sqrt{z L_0}}{\omega_p \xi_e}.
\end{equation}
The radial force acting on an electron has the same sign over sections of length of about $L_n$. If $L_n \gtrsim L_d$, then the electron beam has enough time to respond to fields of each separate wakefield period. Otherwise the oscillating component of the radial force averages out. The condition $L_n = L_d$ is the easiest to meet at the maximum plasma density ($z=L_0$), from which we find
\begin{equation}\label{e11}
    L_0 = \frac{L_d \omega_p \xi_e}{2 \pi c} = N L_d \approx 4\,\text{m},
\end{equation}
where $N \approx 130$ is the electron bunch delay measured in plasma wavelengths. The value of $L_0$ is much longer than the instability growth length (Fig.\,\ref{f3-phases}) and the expected length of the transition region \cite{NIMA-740-197}, so only the period-averaged radial force is of importance in the transition region.

The average of the force (\ref{e6}) always defocuses electrons. The electrons can survive only if  the transition region is shorter than the defocusing distance. Since defocusing is due to the first term in (\ref{e6}), the tolerable transition length is
\begin{equation}\label{e12}
    L_0 \lesssim L_d \exp \left( \frac{\xi_e^2}{4 \sigma_{zb}^2} \right) \approx 4.6\,\text{cm}.
\end{equation}

\begin{figure}[tb]
\bc\includegraphics[width=179bp]{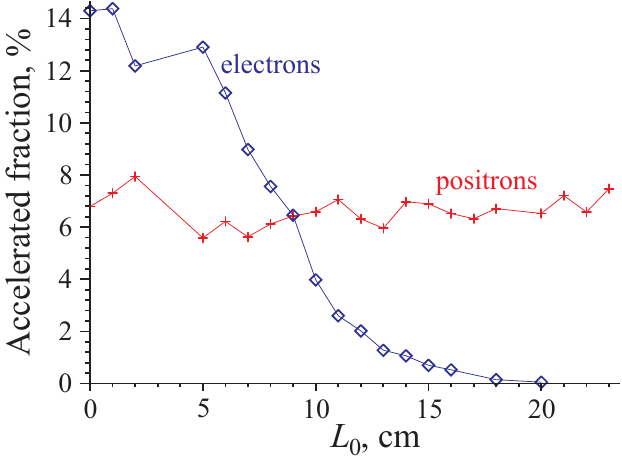} \ec
 \vspace*{-5mm}
\caption{ Accelerated fraction of electron and positron beams versus the length of the transition region $L_0$.}\label{f9-entry}
\end{figure}
Simulations confirm the theoretical predictions for electrons (Fig.\,\ref{f9-entry}). The number of accelerated electrons reduces to zero if the condition (\ref{e12}) is not fulfilled. For positrons the picture is qualitatively different. The average radial force (\ref{e6}) focuses positrons, so there is no negative effect of the density slope on positrons.

\section{Summary}

We demonstrated with simulations that it is possible to inject electron or positron beams along the same line as the proton driver. If the velocity of the injected particles is about, or greater than the phase velocity of the wave at the driver self-modulation stage, then the particles are trapped by the wakefield and kept in the potential wells until the driver beam is fully bunched. After the wakefield amplitude reaches its maximum, the particles trapped at the tail of the driver are efficiently accelerated. The injection delay is of importance, since the wave phase velocity there can exceed the light velocity, which is necessary for high energy gain. The final energy spectrum of accelerated particles is reasonably narrow, with the root mean square energy spread of about 15\% even for injected beams covering several wakefield periods.

If the injected beam is many wakefield periods long, then the trapped charge is limited by beam loading effects. The particles trapped in earlier wave periods hamper trapping in later periods. There is an asymmetry in trapping of electrons and positrons caused by the positive charge of the driver. The initial trapping is better for positrons, but at the acceleration stage a considerable fraction of positrons is lost from the wave. Electrons are not trapped if the plasma density increases smoothly over a too long distance at the plasma entrance. The tolerable density transition is several centimeters long for the baseline parameters of AWAKE experiment. Positrons are not susceptible to the initial density gradient.

The above mechanism of trapping and acceleration could be found in several earlier papers, but was not identified for various reasons. In Refs.\,\cite{PoP18-103101,PAC11-TUOBN5} the attention was paid to the highest energy electrons rather than to energy spectra. In Ref.\,\cite{JPP78-455} the electron bunch delay was optimized for side injection, and electrons were injected at the location where the established phase velocity of the wave was very close to $c$. Correspondingly, a wide energy spectrum was observed. In Ref.\,\cite{JPP78-347} the injected electron beams were as long as the drive beam itself and therefore produced wide energy spectra. In Refs.\,\cite{PRL107-145003,AIP1507-103} the injection delay was shorter than the optimum one, thus resulting in almost no net acceleration.

To conclude, the possibility of the on-axis injection makes proof-of-principle experiments on proton driven plasma wakefield acceleration easier, and this injection scheme can be further optimized for narrower final energy spread.

\acknowledgments

The authors thanks AWAKE collaboration for fruitful discussions. LCODE-based studies are supported by The Russian Science Foundation (grant No. 14-12-00043).  Work of J.V. is partially funded by the Alexander Von Humboldt Foundation. Work of L.A., J.V., R.A.F., and L.O.S. is partially supported by FCT (Portugal) through Grant No. EXPL/FIS-PLA/0834/2012. LCODE simulations are made at Siberian Supercomputer Center SB RAS. OSIRIS simulations are performed under a PRACE award for access to resources on SuperMUC (Leibniz Research Center).


\begin{thebibliography}{38}
\bibitem{AWAKE}
	R.Assmann, R.Bingham, T.Bohl, C.Bracco, et al. (AWAKE Collaboration),
	Plasma Phys. Control. Fusion 56, 084013 (2014).
  \bibitem{RMP81-1229}
   E. Esarey, C. B. Schroeder, and W. P. Leemans,
   Rev. Mod. Phys. \textbf{81}, 1229 (2009).
\bibitem{PRD86-010001}
    J. Beringer et al. (Particle Data Group),
    Phys. Rev. D 86, 010001 (2012).
 \bibitem{NatPhys9-363}
    A.Caldwell, K.Lotov, A.Pukhov, and F.Simon,
    Nature Phys. {\bf 5}, 363 (2009).
 \bibitem{PRST-AB13-041301}
	K.V.Lotov,
	Phys. Rev. ST Accel. Beams {\bf 13}, 041301 (2010).
\bibitem{PAC09-4542}
    R. Assmann, M. Giovannozzi, Y. Papaphilippou, F. Zimmermann, A. Caldwell, G. Xia,
    Generation of short proton bunches in the CERN accelerator complex.
    Proceedings of PAC09, (Vancouver, BC, Canada), p.4542--4544.
 \bibitem{PPCF53-014003}
	A. Caldwell, K. Lotov, A. Pukhov and G. Xia,
	Plasma Phys. Controlled Fusion {\bf 53}, 014003 (2011).
 \bibitem{AIP1229-510}
	G. Xia, A. Caldwell, K. Lotov, A. Pukhov, N. Kumar, W. An, W. Lu, W. B. Mori, C. Joshi, C. Huang, P. Muggli, R. Assmann, F. Zimmermann,
	Update of Proton Driven Plasma Wakefield Acceleration.
	In: Advanced Accelerator Concepts, 14th Workshop, AIP Conference Proceedings, edited by S.H.Gold and G.S.Nusinovich, v.1229, p.510-515 (AIP, 2010).
 \bibitem{IPAC10-4395}
	G.Xia, A. Caldwell,
	Producing Short Proton Bunch For Driving Plasma Wakefield Acceleration.
	Proceedings of IPAC2010 (Kyoto, Japan), p.4395--4397.
 \bibitem{EPAC98-806}
    K.V.Lotov,
    Instability of long driving beams in plasma wakefield accelerators.
    Proc. 6th European Particle Accelerator Conference (Stockholm, 1998), p.806-808.
 \bibitem{PRL104-255003}
	N.Kumar, A.Pukhov, and K.Lotov,
	Phys. Rev. Lett. {\bf 104}, 255003 (2010).
 \bibitem{IPAC13-1179}
	P. Muggli, A. Caldwell, O. Reimann, E. Oz, R. Tarkeshian, C. Bracco, E. Gschwendtner, A. Pardons, K. Lotov, A. Pukhov, M. Wing, S. Mandry, J. Vieira,
	Physics of the AWAKE Project.
	Proceedings of IPAC2013 (Shanghai, China), p.1179-1181.
 \bibitem{TDR}
	AWAKE Collaboration,
	AWAKE Design Report: A Proton-Driven Plasma Wakefield Acceleration Experiment at CERN.
	CERN-SPSC-2013-013; SPSC-TDR-003.
\bibitem{NIMA-740-48}
    C.Bracco, E.Gschwendtner, A.Petrenko, H.Timko, T.Argyropoulos, H.Bartosik, T.Bohl, J.E.Mueller, B.Goddard, M.Meddahi, A.Pardons, E.Shaposhnikova, F.M.Velotti, H.Vincke,
    Nucl. Instr. Meth. A \textbf{740}, 48 (2014).
\bibitem{PRL112-045001}
    Y. Fang, V. E. Yakimenko, M. Babzien, M. Fedurin, K. P. Kusche, R. Malone, J. Vieira, W. B. Mori, and P. Muggli,
    Phys. Rev. Lett. \textbf{112}, 045001 (2014).
\bibitem{NIMA-740-74}
    M.Gross, R.Brinkmann, J.D.Good, F.Gr\"uner, M.Khojoyan, A.Martinez de la Ossa, J.Osterhoff, G.Pathak, C.Schroeder, F.Stephan,
    Nucl. Instr. Meth. A \textbf{740}, 74 (2014).
 \bibitem{AIP1229-467}
	A. V. Petrenko, K. V. Lotov, P. V. Logatchov and A. V. Burdakov,
	The Facility for 500 MeV Plasma Wake-Field Acceleration Experiments at Budker INP.
	In: Advanced Accelerator Concepts, 14th Workshop, AIP Conference Proceedings, edited by S.H.Gold and G.S.Nusinovich, v.1229, p.467-471 (AIP, 2010).
\bibitem{PoP19-063105}
    J. Vieira, Y. Fang, W. B. Mori, L. O. Silva, and P. Muggli,
    Phys. Plasmas \textbf{19}, 063105 (2012).
\bibitem{IPAC13-1235}
    J. Vieira, P. Muggli, O. Reimann, N.C. Lopes, L.O. Silva, E. Adli, S.J. Gessner, M.J. Hogan, S.Z.Li, M.D.Litos, Y. Fang, C. Joshi, K.A. Marsh, W.B. Mori, N. Vafaei,
    Self-modulation and hosing instability of SLAC electron and positron bunches in plasmas.
    Proceedings of IPAC2013 (Shanghai, China), p.1235-1237.
  \bibitem{PRL107-145002}
    C.B.Schroeder, C.Benedetti, E.Esarey, F.J.Gruener, and W. P. Leemans,
    Phys. Rev. Lett. 107 (2011), 145002.
 \bibitem{PRL107-145003}
	A. Pukhov, N. Kumar, T. Tuckmantel, A. Upadhyay, K. Lotov, P. Muggli, V. Khudik, C. Siemon, and G. Shvets,
	Phys. Rev. Lett. 107(14), 145003 (2011).
  \bibitem{PoP19-010703}
    C.B.Schroeder, C.Benedetti, E.Esarey, F.J.Gruener, and W. P. Leemans,
    Phys. Plasmas 19 (2012), 010703.
 \bibitem{JPP78-455}
	K.V.Lotov,
	J. Plasma Phys. 78(4), 455-459 (2012).
 \bibitem{IPAC13-1238}
	K.V. Lotov, A. Sosedkin, E.Mesyats,
	Simulation of Self-modulating Particle Beams in Plasma Wakefield Accelerators.
	Proceedings of IPAC2013 (Shanghai, China), p.1238-1240.
 \bibitem{PoP20-013102}
	K.V.Lotov, A.Pukhov, and A.Caldwell,
	Phys. Plasmas 20(1), 013102 (2013).
\bibitem{NIMA-740-197}
    E.Oz, P.Muggli,
    Nucl. Instr. Meth. A \textbf{740}, 197 (2014).
\bibitem{PoP5-785}
    K.V.Lotov,
    Phys. Plasmas, 1998, v.5, N 3, p.785-791.
\bibitem{LCODE}
 \verb"www.inp.nsk.su/~lotov/lcode".
\bibitem{PRST-AB6-061301}
    K.V.Lotov,
    Phys. Rev. ST Accel. Beams {\bf 6}, 061301 (2003).
 \bibitem{osiris}
	R. A. Fonseca \emph{et al.}, Lect. Notes Comp. Sci. vol. 2331/2002, (Springer Berlin / Heidelberg,(2002).
\bibitem{PRE86-026402}
    C.B.Schroeder, C.Benedetti, E.Esarey, F.J.Gruner, and W.P.Leemans,
    Phys. Rev. E \textbf{86}, 026402 (2012).
\bibitem{PoP20-056704}
    C.B.Schroeder, C.Benedetti, E.Esarey, F.J.Gruner, and W.P.Leemans,
    Phys. Plasmas \textbf{20}, 056704 (2013).
\bibitem{PRL112-205001}
    J. Vieira, W. B. Mori, and P. Muggli,
    Phys. Rev. Lett. \textbf{112}, 205001 (2014).
\bibitem{PoP21-056705}
    J. Vieira, R. A. Fonseca, W. B. Mori, and L. O. Silva,
    Phys. Plasmas \textbf{21}, 056705 (2014).
 \bibitem{PoP18-103101}
	A. Caldwell  and K. V. Lotov,
	Phys. Plasmas {\bf 18}, 103101 (2011).
\bibitem{PP15-429}
    G. K\"uppers, A. Salat, H. K. Wimmel,
    Plasma Physics \textbf{15}, 429-439 (1973).
 \bibitem{PoP3-2753}
    K.V.Lotov,
    Phys. Plasmas, 1996, v.3, N 7, p.2753--2759.
  \bibitem{PAcc20-171}
    P.Chen,
    Part. Accel., v.20 (1987), p.171--182.
  \bibitem{PF30-252}
    R.Keinigs and M.E.Jones
     Phys. Fluids, v.30 (1987), p.252--263.
 \bibitem{PoP20-083119}
    K.V.Lotov,
    Phys. Plasmas 20, 083119 (2013).
 \bibitem{PoP18-024501}
	K.V.Lotov,
	Phys. Plasmas 18(2) (2011) 024501.
 \bibitem{PAC11-TUOBN5}
	G. Xia, A. Caldwell, K. Lotov, A. Pukhov, R. Assmann, F. Zimmermann, C.Huang, J. Vieira, N. Lopes, R.A. Fonseca, L.O.Silva, W. An, C. Joshi, W. Mori, W. Lu, P. Muggli,
	A proposed experimental test of proton-driven plasma wakefield acceleration based on CERN SPS
	Proc. 2011 Particle Accelerator Conference, New York, NY, USA, p.TUOBN5(1-3).
 \bibitem{JPP78-347}
	G. Xia, R. Assmann, R. A. Fonseca, C. Huang, W. Mori, L. O. Silva, J. Vieira, F. Zimmermann and P. Muggli,
	J. Plasma Phys. 78(4), 347-353 (2012).
 \bibitem{AIP1507-103}
	A.Pukhov, T.Tuckmantel, N.Kumar, A.Upadhyay, K.Lotov, V.Khudik, C.Siemon, G.Shvets, P.Muggli, and A.Caldwell,
	Principles of Self-Modulated Proton Driven Plasma Wake Field Acceleration.
	AIP Conf. Proc. 1507, 103-110 (2012).
\end{thebibliography}
\end{document}